\documentclass[12pt,twoside,fleqn]{article}
\usepackage{epsfig}
\usepackage{amssymb}
\begin{document}
\thispagestyle{empty}

\begin{center}
{\large{\bf Lattice effective potential of $(\lambda\Phi^4)_4$:\\[0.2cm] 
nature of the phase transition and bounds on the Higgs mass} }
\end{center}
\vspace{1.0cm}
\begin{center}
{\large P.~Cea$^{1,2,}$\footnote{E-mail: cea@bari.infn.it}, 
L.~Cosmai$^{2,}$\footnote{E-mail: cosmai@bari.infn.it},
M.~Consoli$^{3,}$\footnote{E-mail: consoli@infnct.infn.it}
and R.~Fiore$^{4,5,}$\footnote{E-mail: fiore@cs.infn.it} }\\
\vspace{1.0cm}
$^1$~INFN - Sezione di Bari - via Amendola 173 - I 70126 - Bari - Italy \\
$^2$~Dipartimento di Fisica - Universit\`a di Bari - via Amendola 173 - 
I 70126 Bari - Italy \\
$^3$~INFN - Sezione di Catania - Corso Italia 57 - I 95129 Catania - Italy \\
$^4$~Dipartimento di Fisica - Universit\`a della Calabria - I 87030 Arcavacata di
Rende (Cs) - Italy \\
$^5$~INFN - Gruppo Collegato di Cosenza - I 87030 Arcavacata di
Rende (Cs) - Italy
\end{center}
\vspace{0.8cm}
\begin{center}
{\large {\bf Abstract}}
\end{center}
\vspace{0.3cm}
We present a detailed discussion of Spontaneous Symmetry Breaking (SSB)
in $(\lambda\Phi^4)_4$. In the usual approach, inspired by perturbation
theory, one predicts a second-order phase transition,
the Higgs mass $m_h$, related to the value of the renormalized 4-point
coupling, gets smaller when
increasing the ultraviolet cutoff and this leads to the generally
quoted upper bounds $m_h<$700-900 GeV.
On the other hand, by exploring the structure of the effective potential
in those approximation consistent with `triviality',
where the Higgs mass does not represent a measure of any observable
interaction, SSB does not require an
ultraviolet cutoff, the phase transition
is first-order, such that the massless `Coleman-Weinberg' regime lies in
the broken phase, and one gets only $m_h<$3 TeV from vacuum stability.
To separate out the two alternatives, we present a
precise lattice computation of the slope of the effective
potential in the region of
bare parameters indicated by the Luscher~\&~Weisz 
and Brahm's analysis of the critical line.
Our lattice data strongly support the latter description of SSB.
Indeed, our data cannot be reproduced in perturbation theory, and then they
confirm the existence on the lattice of a remarkable phase of
$(\lambda\Phi^4)_4$  where SSB is generated
through ``dimensional transmutation'', and
show no evidence for residual self-interaction effects of
the shifted ``Higgs'' field $h(x)=\Phi(x)-\langle\Phi\rangle$, in
agreement with ``triviality''.
\newpage
\setcounter{page}{1}

\section{Introduction}

Spontaneous Symmetry Breaking (SSB), induced through a
self-interacting  scalar sector, is the essential ingredient to
generate the mass of the  intermediate vector bosons in the standard
model of electroweak interactions \cite{WS}. However,  despite of the
simplicity and elegance of the Higgs mechanism \cite{higgs}, it is
believed that the generally accepted ``triviality'' 
\cite{aizen,froh,sokal,latt,luscher87,glimm,call,book} of $(\lambda\Phi^4)_4$
implies the scalar sector of the standard model to be just an
effective theory, valid up to some cutoff scale.   Without a cutoff,
the argument goes, there would be no scalar self-interactions,  and
without them  there would be no symmetry breaking. This point of view
also leads to upper bounds on the Higgs mass \cite{call,neu,lang}.

Recently \cite{resolution,zeit}, it has been pointed out, on the 
basis of very general arguments, that SSB is not incompatible with
``triviality''.  Indeed, the most general condition for a trivial 
scattering matrix requires {\em all} interaction  effects to be
reabsorbed into a set of Green's  functions expressible in terms of
the first two moments of a Gaussian functional distribution. In this
situation,  where the simple Hartree-Fock-Bogolubov approximation 
becomes effectively  exact in the continuum limit, one can
meaningfully consider a non-zero vacuum expectation value (VEV) of the
field,  $\langle\Phi\rangle$, in connection with non-interacting,
quasi-particle excitations in the broken  symmetry phase. In this
sense, a ``trivial'' theory can still be useful to  determine the
vacuum of the theory and as such a convenient frame for the  gauge
fields preserving a perturbatively weak interaction at high energies.
This picture, while reconciling the strong evidences for
``triviality'' with those for a non-trivial effective potential
\cite{cian,return,cast,munoz,bran,con,new,iban,rit,pg,rit2},  also
clarifies the meaning  of some explicit studies of ``triviality'' in
the {\em broken} symmetry phase ($\langle\Phi\rangle \ne 0$)
\cite{broken}, all pointing out the  absence of {\em observable}
interactions. Let us briefly recapitulate the simple logical steps of
Ref.~\cite{resolution,zeit} leading to this conclusion.

The effective potential determines the zero-momentum 1PI vertices 
\cite{CW} and reflects any non trivial dynamics in the zero-momentum
sector of the theory. This remark  immediately suggests the relevance
of studying the {\em massless} version of $(\lambda\Phi^4)_4$
theories. In fact, on one hand,  just the situation of a vanishing
mass-gap in the symmetric phase, preventing  to deduce the uniqueness
of the vacuum from the basic results of quantum field theory (see
Ref.~\cite{glimm}, chapts. 16 and 17), opens the possibility of SSB.
On the other hand,  since for the massless theory the zero-momentum
($p_{\mu}=0$) is a physical on-shell point, a non-trivial effective
potential implies a non-trivial scattering matrix, at least in that
limiting and unobservable region of the 4-momentum space, in agreement
with the independent evidence for ``non-triviality''  of Pedersen,
Segal and Zhou \cite{ped}. As pointed out in 
Ref.~\cite{resolution,zeit}, in fact,  this (admittedly almost
ignored) result does not imply, by itself,  the presence of {\em
observable} scattering processes since all non-trivial interaction
effects of the symmetric phase may become {\em unobservable}, being
fully  reabsorbed into a change of the vacuum structure and in the
mass of the excitations of the broken symmetry vacuum. In this way,
one can reconcile non-trivial SSB with ``triviality'' and, in the
simplest case of the one-component theory with a discrete symmetry
$\Phi \to -\Phi$, the physical  excitation spectrum contains only
massive free particles. Furthermore, by relating the equivalent
computations in the symmetric phase of Coleman and Weinberg \cite{CW}
with those in the broken phase by Jackiw \cite{roman} the  requirement
of ``triviality'', i.e. the  condition of a free shifted `Higgs' field
$h(x)=\Phi(x)-\langle\Phi\rangle$, dictates the form of the  effective
potential which, close to the continuum limit, has to reduce to the
simple expression corresponding to the sum of a classical background
and of the zero-point energy of a massive free field. Indeed, 
massless $(\lambda\Phi^4)_4$, in all approximations consistent with 
``triviality'', i.e.  when $h(x)$   is effectively governed by a
quadratic Hamiltonian (one-loop potential,  Gaussian approximation,
postgaussian calculations, see in particular Ref.~\cite{rit2}, where
the  Higgs propagator $G(x,y)$ is properly optimized at each value of
$\langle\Phi\rangle$, by solving the corresponding non-perturbative
gap-equation) provides the same structure for the effective potential
and, close to the continuum limit, one finds the simple expression
\cite{con,new,resolution,zeit} ($\phi_B=\langle\Phi\rangle$ denotes
the bare, cutoff-dependent  vacuum field and $V_{\mathrm{triv}}$ is a
short-hand notation to denote the effective potential in those
approximations consistent with ``triviality'')
\begin{equation}
\label{eq:1}
V_{\mathrm{triv}}(\phi_B)  =  \frac{\tilde\lambda}{4} \phi^4_B + 
\frac{\omega^4(\phi_B)}{64\pi^2}  \left( \ln
\frac{\omega^2(\phi_B)}{\Lambda^2} -  \frac{1}{2} \right) \,.  
\end{equation}
In Eq.~(\ref{eq:1}) $\Lambda$ denotes the Euclidean ultraviolet
cutoff,  $\omega^2(\phi_B)=3\tilde\lambda\phi^2_B$ is the
$\phi_B-$dependent mass  squared of the shifted field  and
$\tilde\lambda$ is finitely proportional to the bare coupling
$\lambda_0$ entering the  bare Lagrangian density
($\tilde\lambda=\lambda_0$ at one-loop, 
$\tilde\lambda=(2/3)\lambda_0$ in the Gaussian approximation
and so on). Eq.~(\ref{eq:1}) provides the most general form of the
effective potential in those approximations consistent with
``triviality'' and leads to SSB. Indeed,  the absolute minimum
condition occurs at $\phi_B=\pm v_B\neq 0$ where
\begin{equation}
\label{eq:2}
m^2_h=3\tilde\lambda v^2_B=
\Lambda^2\exp \left[ -\frac{16\pi^2}{9\tilde\lambda} \right] 
\end{equation}
and one finds
\begin{equation}
\label{eq:3}
W=V_{\mathrm{triv}}(\pm v_B)=-\frac{m^4_h}{128\pi^2} <0
\end{equation}
so that, by using Eq.~(\ref{eq:2}), 
Eq.~(\ref{eq:1}) can be rewritten as
\begin{equation}
\label{eq:4}
V_{\mathrm{triv}}(\phi_B)=\frac{9\tilde\lambda^2\phi^4_B}{64\pi^2}
\left( \ln \frac{\phi^2_B}{v^2_B} - \frac{1}{2} \right) \,.
\end{equation}
As discussed in Ref.~\cite{con,new,resolution,zeit,rit,rit2}, a
non-perturbative renormalization of the ``trivial'' effective
potential is possible by imposing the requirement of
cutoff-independence for the ground state energy density  $W$ in
Eq.~(\ref{eq:3}) and for the physical correlation length  $\xi_h\sim
1/m_h$ in the broken phase. In this case, from
\begin{equation}
\label{eq:5}
\Lambda \frac{dm_h}{d\Lambda}=  \left( \Lambda 
\frac{\partial}{\partial\Lambda}+ \beta(\tilde\lambda)
\frac{\partial}{\partial\tilde\lambda} \right) m_h=0
\end{equation}
one deduces
\begin{equation}
\label{eq:6}
\beta(\tilde\lambda)=\Lambda \frac{d\tilde\lambda}{d\Lambda}
=-\frac{9\tilde\lambda^2}{8\pi^2}<0
\end{equation}
which implies that both $\lambda_0$ and $\tilde\lambda$ vanish in the
continuum limit $\Lambda\to \infty$, $m_h={\mathit{fixed}}$ according to
\begin{equation}
\label{eq:7}
\lambda_0\sim 3\tilde\lambda=\frac{m^2_h}{v^2_B}= 
\frac{8\pi^2}{3\ln(\Lambda/m_h) } \to 0.
\end{equation}
Notice that the non-perturbative $\beta$-function obtained from
``triviality'' is very different from the perturbative
$\beta$-function
\begin{equation}
\label{eq:8}
\beta_{\mathrm{pert}}(\lambda_0)=\Lambda{{d\lambda_0}\over{d\Lambda}}=
{{9\lambda^2_0}\over{8\pi^2}}- {{51\lambda^3_0}\over{64\pi^4}}+
O(\lambda^4_0)
\end{equation}
deduced from the cutoff-independence  of the {\em renormalized}
coupling $\lambda_R= \lambda_R(\mu^2)$, as computed in a weak coupling
expansion \cite{beta} in powers of the bare coupling $\lambda_0$ at
some non-zero external momenta $p^2=\mu^2$
\begin{equation}
\label{eq:9}
\Lambda{{d\lambda_R}\over{d\Lambda}}= (\Lambda{{\partial
}\over{\partial\Lambda}}+\beta_{\mathrm{pert}}(\lambda_0)
{{\partial}\over{\partial\lambda_0}})\lambda_R=0.
\end{equation}
In particular, by relying on a perturbative evaluation of the higher
order effects one would deduce \cite{CW} that the one-loop minimum is 
changed into a false vacuum by the genuine $h-$field self-interactions
and that SSB does not occur for the massless theory.  In this
framework, and quite independently of our results, one should realize
that in the ``trivial'' $(\lambda\Phi^4)_4$ theories the validity of
the perturbative relations is, at best, unclear \cite{note}. Indeed,
perturbation theory, being based on the concept of a
cutoff-independent and non-vanishing renormalized coupling at non-zero
external momenta, predicts the existence of observable scattering
processes that cannot be there if ``triviality'' is true. In this
sense, the unphysical features of the perturbative $\beta-$function
\cite{beta} are just a consequence \cite{trivpert} of the basic
assumption behind the perturbative approach -- the attempt of defining
a continuum limit in the presence of  residual interaction effects
which, at any finite order, cannot be reabsorbed into the vacuum
structure and the particle mass. Therefore, only abandoning from the
start the vain attempt of defining an interacting theory, can a 
meaningful continuum limit be obtained and only approximations to the
effective potential consistent with the non-interacting nature of the
field $h(x)$ are reliable (for more details on the general class of
these consistent approximations see Ref.~\cite{zeit}).

As discussed in Ref.~\cite{con,resolution,zeit,new,rit,rit2},   all
approximations consistent with the structure in
Eqs.~(\ref{eq:1},\ref{eq:4}), although  providing different
$\tilde\lambda$ in their bare forms, are {\em equivalent}. Indeed,
they lead to the same form when expressed in terms of the physical
{\em renormalized} vacuum field $\phi_R$. The underlying  rationale
for the introduction of $\phi_R$ is the following.  Naively, one would
plot the effective potential (\ref{eq:1},\ref{eq:4}) in terms of the
bare field $\phi_B$ at fixed $\Lambda$. This attempt, however, becomes
more and more difficult when increasing $\Lambda$  since in the
continuum limit,  when $\Lambda\to\infty$ and $\tilde\lambda\to 0$, 
the slope of the effective potential in terms of $\phi_B$ becomes
infinitesimal. Indeed,  to produce  a finite change in
$V_{\mathrm{triv}}$ from $0$ to $W$ requires  to reach the values $\pm
v_B$ and these are located at an infinite distance in units of $m_h$.
Thus, in this situation the plot of the effective potential is
crucially dependent on the magnitude of the ultraviolet cutoff.
However, the effective potential itself {\em is} a cutoff-independent
quantity and one may naturally consider the question of making this
invariance {\em manifest}. To do this, let us consider the 
Renormalization Group (RG) equation for the effective potential and
determine the integral curves
\begin{equation}
\label{eq:10}
\tilde\lambda=\tilde\lambda(\Lambda)  \;,
\end{equation}
\begin{equation}
\label{eq:11}
\phi_B=\phi_B(\Lambda)                  
\end{equation}
along which $V_{\mathrm{triv}}$ in Eqs.~(\ref{eq:1},\ref{eq:4}) is
invariant. This amounts to solve the partial differential equation
\begin{equation}
\label{eq:12}
(\Lambda{{\partial }\over{\partial\Lambda}}+\beta(\tilde\lambda)
{{\partial}\over{\partial\tilde\lambda}}+ \Lambda{{d \phi_B}\over{d
\Lambda}} {{\partial }\over{\partial
\phi_B}})V_{\mathrm{triv}}(\Lambda,\tilde\lambda,\phi_B)=0 \,.
\end{equation}
By using Eq.~(\ref{eq:6}), obtained from the cutoff-independence of
the ground state energy at the minima $\phi_B=\pm v_B$ where the
partial derivative with respect to $\phi_B$ in 
Eq.~(\ref{eq:12}) vanishes
identically, we find~\cite{cast,bran,con,resolution,rit}
\begin{equation}
\label{eq:13}
\Lambda{{d
\phi_B}\over{d\Lambda}}={{9\tilde\lambda}\over{16\pi^2}}\phi_B
\end{equation}
which indeed confirms that the product $\tilde\lambda\phi^2_B$ is
invariant along a given integral curve as dictated by the
RG-invariance of Eq.~(\ref{eq:2}). Thus, one finds $\phi^2_B\sim
1/\tilde\lambda$ and the question  naturally arises of finding the
proper normalization of the vacuum field in units of the natural
cutoff-independent scale $m_h$ associated with the absolute minimum of
the effective potential. To this end, let us consider the general
structure in Eq.~(\ref{eq:4}) and introduce the cutoff-independent
combination $\phi_R$ such that
\begin{equation}
\label{eq:14}
\phi^2_R={{3\tilde{\lambda}\phi^2_B}\over{8\pi^2 X}}
\equiv{{\phi^2_B}\over{ Z_{\phi}}} \,,
\end{equation}
$X$ being an arbitrary, positive $\tilde\lambda$-independent number.
In this way 
\begin{equation}
\label{eq:15}
V_{\mathrm{triv}}(\phi_B)= V_{\mathrm{triv}}(\phi_R)=\pi^2X^2\phi^4_R
(\ln{{\phi^2_R}\over{v^2_R}}-{{1}\over{2}})
\end{equation}
so that, from the value at $\phi_B=\pm v_B$,
%
\begin{equation}
\label{eq:16}
W  =  V_{\mathrm{triv}}(\pm v_B)=V_{\mathrm{triv}}(\pm v_R)
 =  -{{1}\over{2}}\pi^2X^2v^4_R=-{{m^4_h}\over{128\pi^2}}
\end{equation}
we find 
\begin{equation}
\label{eq:17}
m^2_h=8\pi^2Xv^2_R \,.
\end{equation}
However,  when considering the second derivative of the effective
potential with respect to $\phi_R$ at $\pm v_R$ 
\begin{equation}
\label{eq:18}
\left. \frac{d^2V_{\mathrm{triv}}(\phi_R)}{d\phi^2_R}
\right|_{\phi_R=\pm v_R}=8\pi^2 X^2 v^2_R
\end{equation}
and depending on the value of $X$, the quadratic shape in terms of the
rescaled field $\phi_R$ will not agree with the Higgs mass $m_h$
unless
\begin{equation}
\label{eq:19}
X=1.
\end{equation}
In this case, namely for
\begin{equation}
\label{eq:20}
V_{\mathrm{triv}}(\phi_R)=\pi^2\phi^4_R
(\ln{{\phi^2_R}\over{v^2_R}}-{{1}\over{2}}),
\end{equation}
\begin{equation}
\label{eq:21}
m^2_h=8\pi^2v^2_R,
\end{equation}
when the effective potential at its minima is locally equivalent to a
harmonic potential parametrized in terms of the {\em physical} Higgs
mass, the continuum theory associated with $X=1$  is completely
indistinguishable from a free-field theory. 
Eqs.~(\ref{eq:20}, \ref{eq:21}),  discovered in all known approximations
consistent with ``triviality'', should be considered {\em  exact}
(strictly speaking the exact effective potential is the ``convex
hull'' of Eq.~(\ref{eq:20})~\cite{book}).

The only unconventional ingredient of the analysis is the  asymmetric 
rescaling of vacuum field and fluctuation implying $Z_{\phi}\neq Z_h$,
since in any approximation consistent with ``triviality'' there is no
non-trivial  rescaling of $h(x)$. However, the introduction of
$Z_{\phi}$ is extremely natural when considering the quantization of
the classical Lagrangian
\begin{displaymath}
L={{1}\over{2}}(\partial\Phi)^2
-{{1}\over{2}}r_0\Phi^2
-{{1}\over{4}}\lambda_0\Phi^4 \,.
\end{displaymath}
In fact,  as it is well known from statistical mechanics, a consistent
quantization requires to single out preliminary the zero-momentum
mode $\phi_B$
\begin{displaymath}
\Phi(x)=\phi_B+h(x)
\end{displaymath}
(determined from the condition $\int d^4x~h(x)=0$), whose essentially
classical nature reflects, however, the fundamental quantum phenomenon
of Bose condensation. Therefore, beyond perturbation theory, the
renormalization of the term containing the field derivatives, defining
$Z_h$, is quite unrelated to $Z_{\phi}$, defined from the scaling
properties of the effective potential, and, in those approximations
consistent with ``triviality'', e.g.  preserving $Z_h=1$ identically
as at one-loop or in the Gaussian  approximation, one finds
$Z_{\phi}\sim 1/\lambda_0$. The structure $Z_{\phi}\neq Z_h$, allowed
by the Lorentz-invariant meaning of the field decomposition into
$p_{\mu}=0$ and $p_{\mu}\neq 0$ components \cite{resolution,zeit}, is
completely consistent with the rigorous indications of quantum field
theory. In fact, SSB in the cutoff theory  (see Ref.~\cite{book},
Sect.15), while imposing the {\em finiteness} of $Z_h$,  requires that
the bare vacuum field $v_B$ and the Higgs mass $m_h$ do not scale
uniformly and one finds
\begin{equation}
\label{eq:22}
{{m^2_h}\over{v^2_B}} \to 0
\end{equation}
in the continuum limit where the lattice spacing  $a\sim 1/\Lambda\to
0$ in agreement with Eq.~(\ref{eq:7}). Equation~(\ref{eq:22}), by
itself, represents a perfectly acceptable result. Indeed, there is no
reason, in  principle, why the bare vacuum field,   defined at the
lattice level, should be finite. One is actually faced with the
situation where the vacuum energy $W=V_{\mathrm{eff}}(\pm v_B)$  (related
to the critical temperature $T_c$ at which the symmetry is restored
\cite{hajj}) is finitely related to $m^2_h$ even though $v^2_B$ itself
is not. A familiar example of this situation is gluon condensation in
QCD where the vacuum energy, as measured with respect to the
perturbative ground state, is finitely related to the particular
combination of bare  quantities $g^2_B\langle G^2_B \rangle$. In the
continuum limit where  $g^2_B\to 0$ the bare expectation value
$\langle G^2_B \rangle$ diverges in units of the physical scale of the
QCD vacuum as defined, for instance, from the string tension or a
glueball mass. The same is true in $(\lambda\Phi^4)_4$ where the bare
vacuum field $v_B$ does not remain finite in units of $m_h$, the
physical scale setting the correlation length of the  spontaneously
broken phase. However,  despite of this illuminating analogy, pointing
out the very general nature of the problem,  the trend in
Eq.~(\ref{eq:22}) is usually considered as an indication for the
inconsistency of SSB in the continuum limit. The reason for this
conclusion, most likely, derives from the operatorial meaning given to
the field rescaling. Namely, the perturbative assumption of a single
renormalization constant $Z=Z_{\phi}=Z_h$, to account for the
cutoff-dependence both of the vacuum field and of the residue of the 
shifted field propagator, amounts to introduce a  {\em renormalized
field operator}
\begin{equation}
\label{eq:23}
\Phi(x)=\sqrt{Z}~\Phi_R(x) \,,
\end{equation}
This relation is a consistent shorthand for expressing the {\em wave
function} - renormalization in a theory allowing an asymptotic Fock
representation, but, beyond perturbation theory, has no justification 
in the presence of SSB. In fact, in this case, it overlooks that the
shifted field $h(x)$,  the only allowing for a particle
interpretation,  is not defined before fixing the vacuum and that the
Lehmann  spectral-decomposition argument ($0<Z\leq 1$), valid for a
field with  vanishing VEV, constrains only the value of $Z=Z_h$. 
Notice that, quite independently of our results, the possibility of a
different rescaling for the vacuum field and the fluctuations is 
somewhat implicit in the conclusions of the authors of
Ref.~\cite{book} (see their footnote at page 401: "This is reminiscent
of the standard procedure in the central limit theorem for independent
random variables with a nonzero mean: we must subtract a mean of order
$n$ before applying the rescaling $n^{-1/2}$ to the fluctuation
fields"). 

Equation~(\ref{eq:22}), supported as well by the results of lattice
calculations (see Ref.~\cite{lang} and Sect.3),  represents a basic
condition to define the continuum limit of SSB in $(\lambda\Phi^4)_4$.
At the same time, by requiring $m^2_h$ to be cutoff independent,
Eq.~(\ref{eq:7}) implies
\begin{equation}
\label{eq:24}
\lambda_0 \sim \tilde\lambda \to 0
\end{equation}
for $\Lambda\to\infty$ and, thus,  one recovers, with very different
techniques, the consistency condition usually denoted as ``asymptotic 
freedom'' \cite{froh,call}. Obviously, in the trivial
$(\lambda\Phi^4)_4$  theory, where no observable scattering process
can survive in the continuum  limit, the notion of asymptotic freedom
is very different from QCD and has nothing to do with the existence of
a renormalized coupling constant $\lambda_R(\mu^2)$ at non vanishing
external momenta. Only at zero momentum a non-trivial dynamics is
possible and this unobservable effect  is fully reabsorbed in the
vacuum structure and in the particle mass of the field $h(x)$.
Thus ``triviality'' means that the finite-momentum modes are
free-field like. The physics in the 
SSB vacuum has free particles, and $m_h$ is the mass of any physical 
(on-shell) particle with any 3-momentum one wishes, including zero.  
It is true that there is a subtlety about zero 4-momentum, but this affects 
only off-shell particles, and since the theory is free the experimenter 
has no way to make off-shell particles.

The remarkable consistency of this theoretical framework suggests to
look for additional tests of the structure of the effective potential
in  Eqs.~(\ref{eq:1},\ref{eq:4}) by comparing with precise lattice
simulations of the massless regime. To this end, a few general remarks
are needed. While substantial experience has been already gained in
the numerical analysis of lattice field theories, the reliability of
statements about their continuum limits is still open to questions. 
In fact, the equivalence of different procedures to get those limits
is not yet controlled by standard theoretical methods. These imply a
shift of the problem from the domain of bare computation, including
the evaluation of numerical errors and/or approximations, to that of
interpretation, exploiting the connection of the numerical results 
with the formal theory or with suitable models of the continuum limit.
For instance, in all lattice simulations  performed so far, the
perturbative relation $Z_{\phi}\equiv Z_h$ has been assumed to define, 
from the average bare field measured on the lattice,  a renormalized
vacuum  field that, in the O(4)-symmetric case, is then related to the
Fermi constant.  In this case, since the lattice  data definitely
support the trend in Eq.~(\ref{eq:22}) and provide trivially free
shifted fields well consistently with $Z_h=1$  (see Ref.~\cite{lang}),
one frequently deduce upper limits on the Higgs mass which would not
exist otherwise. Therefore, a fully  {\em model-independent} approach
is needed to check which relations are actually valid outside the
perturbative domain.
To this end, the lattice approach to quantum field theories can be
very useful. Indeed the lattice offers us the unique opportunity to
study a quantum field theory with an ultraviolet cutoff (the inverse
of the lattice spacing) by means of non-perturbative methods.

An important progress in this direction  has been performed in
Ref.~\cite{andro}. There, the lattice theory defined by the Euclidean
action
%
\begin{equation}
\label{eq:25}
S = \sum_x \left[ \frac{1}{2} \sum_{\hat{\mu}}(\Phi(x+\hat{\mu}) -
\Phi(x))^2  +  \frac{r_0}{2} \Phi^2(x) 
+  \frac{\lambda_0}{4} \Phi^4(x) - J \Phi(x) \right]
\end{equation}
($x$ denotes a generic lattice site and, unless otherwise stated, 
lattice units are understood) was used to compute the VEV of the bare
scalar field $\Phi(x)$
in the presence of an ``external source'' whose strength $J(x)=J$  is
$x$-independent
\begin{equation}
\label{eq:26}
\langle \Phi \rangle _J = 
\left\langle \frac{1}{L^4} \sum_x  \Phi(x) \right\rangle_J=
\phi_B(J) \,,
\end{equation}
where L is the linear dimension of the lattice.  
Determining $\phi_B(J)$ at several $J$-values is
equivalent \cite{huang,huang2,call2} to inverting  the relation 
\begin{equation}
\label{eq:27}
J=J(\phi_B)={{dV_{\mathrm{eff}}}\over{d\phi_B}}
\end{equation}
and starting from the action in Eq.~(\ref{eq:25}), the effective
potential  can be {\em rigorously} defined for the lattice theory up
to an arbitrary integration constant. In this framework, the
occurrence of SSB is determined  by exploring (for $J\neq0$) the
properties of the function 
\begin{equation}
\label{eq:28}
\phi_B(J)=-\phi_B(-J)
\end{equation}
in connection with its behaviour in the limit of zero external source
\begin{equation}
\label{eq:29}
\lim_{J\to 0^{\pm}}~\phi_B(J)=\pm v_B \neq 0
\end{equation}        
over a suitable range of the bare parameters  $r_0,\lambda_0$
appearing in Eq.~(\ref{eq:25}). The existence of non-vanishing
solutions of Eq.~(\ref{eq:29}) is associated with non-trivial extrema
of the effective potential at $\phi_B=\pm v_B$ whose energy density is
{\em lower} than (or equal to) the corresponding value in the
simmetric phase $\phi_B=0$. In fact, through the Legendre transform
formalism, the resulting effective potential is  convex downward and
represents the convex hull \cite{book} of the more familiar  ``double
well'' effective potential (for more details see 
Ref.~\cite{rit,convex}). 

The analysis of Ref.~\cite{andro} was performed for weak bare coupling
$\lambda_0/\pi^2<<1$,  to approach the ``triviality''
continuum limit in Eq.~(\ref{eq:24}), and exploring the dependence on
the bare mass $r_0$ to provide an operative definition of the massless
regime. Here, some remarks are needed. The massless theory is defined
\cite{CW} from  the condition
\begin{equation}
\label{eq:30}
\left. \frac{d^2V_{\rm{eff}}}{d\phi^2_B} \right|_{\phi_B=0}=0 \;,
\end{equation}
ensuring that the theory has no physical mass scale in its symmetric
phase $\phi_B=0$. Thus, from Eq.~(\ref{eq:27}),  one should explore,
in principle, the shape of $J=J(\phi_B)$ around $J=0$. However, in any
finite lattice \cite{call2}, the basic inadequacy of the finite volume
calculation shows up in the occurrence of large finite size effects at
very small values of $|J|$.
This suggests that a reliable estimate
of the massless regime requires to evaluate the response of the 
system at non-zero $J$ and then to {\em extrapolate} the lattice data
toward $J=0^{\pm}$ with some  analytic form. To this end, in
Ref.~\cite{andro} Eq.~(\ref{eq:30}) was replaced by a trial form for
the source consistent with the ``triviality'' structure in 
Eqs.~(\ref{eq:1},\ref{eq:4}) but allowing for an explicit scale
breaking parameter $\beta$ \cite{zeit}:
\begin{equation}
\label{eq:31}
J(\phi_B)=\alpha\phi^3_B\ln(\phi^2_B)+\beta\phi_B+\gamma\phi^3_B
\end{equation}
(for $\beta=0$ Eqs.~(\ref{eq:27},\ref{eq:31}) imply 
$\alpha=(9\tilde\lambda^2)/(16\pi^2)$ and
$\gamma=(9\tilde\lambda^2)/(16\pi^2)\ln(1/v^2_B)$ in the
case of the effective potential in Eqs.~(\ref{eq:1},\ref{eq:4})).

The massless regime at each $\lambda_0$ was operatively defined from
the maximum value  of the bare mass $r_0$ where the 3-parameter fit
$(\alpha,\beta,\gamma)$  to the lattice data (for $|J|\geq 0.05$) was
giving exactly the same $\chi^2$ of the 2-parameter fit
$(\alpha,\beta=0,\gamma)$. Finally, the numerical results of
Ref.~\cite{andro}  were also fitted with analytical forms allowing for
the presence of perturbative corrections. As a matter of fact,
Eqs.~(\ref{eq:1},\ref{eq:4}) agree remarkably well with the lattice
results, while the ``pro-forma'' perturbative leading-log improvement
fails to reproduce the Monte Carlo data, thus providing a definite
numerical support for the coexistence of SSB and ``triviality''
proposed in  Ref.~\cite{resolution,zeit}. 

A possible objection to this procedure is that the operative
definition of the massless theory is not fully model-independent
since, by using Eq.~(\ref{eq:31}), one essentially checks the
self-consistency of the procedure while a definitive test of the
structure of the effective potential  requires an {\em a priori}
estimate of $r_c$ as, for instance, determined from the lattice data 
of other groups.

After this general introduction,  the aim of the present paper is just
to provide  this more refined lattice test of the ``triviality''
structure in Eqs.~(\ref{eq:1},\ref{eq:4}). Our analysis will be
presented in Sect.~2 while Sect.~3 will contain our conclusions and a
discussion of the more general  consequences of our results.

\section{Lattice simulation of massless $(\lambda\Phi^4)_4$}

The starting point for the analysis of the massless theory is the 
Schwinger-Dyson equation in the symmetric phase which determines the
pole of the scalar propagator from the 1PI self-energy
\begin{equation}
\label{eq:32}
m^2=r_0+ \Sigma(p^2=m^2)
\end{equation}
and leads to the zero-mass condition for the ``critical'' value of the
bare mass
\begin{equation}
\label{eq:33}
r_c=-\Sigma(p^2=0) \,.
\end{equation}
Depending on the adopted regularization scheme, Eq.~(\ref{eq:33})
disposes of the bare  mass in the classical lagrangian as a
counterterm for the quantum theory and,  in this situation, the theory
does not possess any physical scale in its symmetric phase
$\langle\Phi\rangle=0$. In dimensional regularization  $r_c=0$ and the
scale invariance of the classical theory is preserved up to
logarithmic terms. Concerning the lattice theory some remarks are
needed.

We stress that the definition of the massless theory is independent on
any assumption about the nature of the phase transition in
$(\lambda\Phi^4)_4$. More precisely, from the general  structure of
the effective potential in  Eqs.~(\ref{eq:1},\ref{eq:4}), we deduce
that the system is already in the broken phase at $m^2=0$ in contrast
with the leading-log improvement of the one loop result \cite{CW}. To
clarify this point it is worthwhile to recall some rigorous results.
In the symmetric phase $\langle\Phi\rangle=0$,  the existence of the
$(\lambda\Phi^4)_4$ {\em critical point} can be established for the
lattice theory (see chapt.17 of Ref.~\cite{glimm}). Namely, for any
$\lambda_0 > 0$, a critical value $r_c=r_c(\lambda_0)$ exists, such
that the quantity $m(r_0,\lambda_0)$ defined as 
\begin{equation}
\label{eq:34}
m(r_0,\lambda_0)= -\lim_{|x-y|\to\infty}~{
{\ln\langle\Phi(x)\Phi(y)\rangle} \over{|x-y|} }
\end{equation}
is a continuous, monotonically decreasing, non-negative function for
$r_0$ approaching $r_c$ from above and one has $m(r_0,\lambda_0)=0$
for $r_0= r_c(\lambda_0)$.   For $r_0 > r_c(\lambda_0)$, 
$m(r_0,\lambda_0)$ is the energy of the lowest nonvacuum state in the
symmetric phase $\langle\Phi\rangle=0$.

The basic problem for the analysis of the phase transition  in the
{\em cutoff} theory concerns the relation between $r_c(\lambda_0)$ and
the value marking the onset of SSB, say $r_s(\lambda_0)$, defined as
the supremum of the values of $r_0$ at which Eq.~(\ref{eq:29})
possesses non-vanishing solutions for a given $\lambda_0$. It should
be obvious that, in general, $r_s$ and $r_c$ correspond to  basically
different quantities. Indeed, $r_c$ defines the limiting situation
where there is no gap for the first excited state in the symmetric
phase  $\langle\Phi\rangle=0$, while  $r_s$ can only be determined
from a stability analysis after computing the relative magnitude of
the energy density in the symmetric and non-symmetric vacua.
A widely accepted point of view, based on the
picture of a second-order Ginzburg-Landau  phase transition with
perturbative quantum corrections, is that, indeed, the system is in
the broken phase at $r_0=r_c-M^2$ for any $M^2>0$, thus implying
$r_s=r_c$. In this case, $-M^2$ represents the `negative renormalized
mass squared' frequently used to describe SSB in pure $\lambda\Phi^4$
and related to the second derivative of the effective potential at the
origin in field space $\phi_B=0$. However, in this picture the massless theory
at $r_0=r_c$ still belongs to the symmetric phase
thus implying that for
\[
r_0\to r_c= r_c(\lambda_0)
\]
(from above) one finds a `continuum limit' of the theory at that
particular value of the bare coupling. Indeed, by using
$m(r_0,\lambda_0)$ in Eq.~(\ref{eq:34})
 to set the physical scale of the theory and
 approaching the critical point, the correlation length would diverge
in units of the lattice spacing.
\par This conclusion, however, ignores
the possibility that
\[
r_c(\lambda_0)<r_s(\lambda_0)
\]
so that the massless regime lies in the broken phase. In this case,
there is no continuum limit at finite bare coupling, since
the physical correlation length
of the theory is not defined through Eq.~(\ref{eq:34})
but has to be determined from the mass of
the shifted field after subtracting
out the field vacuum expectation value.
Thus, understanding the continuum limit of $(\lambda\Phi^4)_4$ requires
to study the effective potential of the theory by exploring
the stability of the symmetric phase for $r_0=r_c(\lambda_0)+\epsilon$
at any $\epsilon > 0$.

In general, the effective
potential is not a finite order polynomial function of the vacuum
field  $\phi_B$ and (see chapt.1 of Ref.~\cite{book} and chapt.4 of
Ref.~\cite{toledo}) one should consider the more general situation
\begin{equation}
\label{eq:35}
V_{\mathrm{eff}}(\phi_B)= {{1}\over{2}}a\phi^2_B +{{1}\over{4}}b\phi^4_B
+{{1}\over{6}}c\phi^6_B + \dots
\end{equation}
where $a,b,c,\ldots$ depend on the bare parameters ($r_0,\lambda_0$)
so that there are several patterns in the phase diagram. In
particular,  if the coefficient $b$ can become negative, even though
$a>0$, there is a first order phase transition and the system dives in
the broken phase passing  through a degenerate configuration where 
$V_{\mathrm{eff}}(\pm v_B)=V_{\mathrm{eff}}(0)$. A remarkable example of
this situation was provided in Ref.~\cite{halperin}. There,  the
superconducting phase transition was predicted to be (weakly) first
order, because of the effects of the intrinsic fluctuating magnetic
field which induce a negative fourth order coefficient in the free
energy when the coefficient of the quadratic term is still positive.
Unfortunately, the  predicted effect is too small to be measured but,
conceptually, is extremely relevant.

Thus, on general grounds, one may consider the possibility that 
$r_c<r_s$ even though for $r_0 > r_c$ the symmetric phase of the
lattice $(\lambda\Phi^4)_4$ has still a mass gap and an exponential
decay of the two-point correlation function. The subtlety is that the
general theorem 16.1.1 of  Ref.~\cite{glimm}, concerning the
possibility of deducing the uniqueness of the vacuum in the presence
of a non-vanishing mass gap in the symmetric phase, holds for the {\em
continuum} theory. Namely, by introducing the variable 
$t=\ln(a_0/a)$, $a_0$ being a fixed length scale, and defining
the continuum limit as a suitable path in the space of the bare
parameters $r_0=r_0(t)$, $\lambda_0=\lambda_0(t)$  (with
$\lambda_0(t)\sim 1/t$ according to Eqs.~(\ref{eq:7},\ref{eq:24}))
only paths leading to a vanishing mass gap in the symmetric phase,
i.e. for which
\begin{equation}
\label{eq:36}
\lim_{t\to \infty} ~m(r_0(t),\lambda_0(t))=0
\end{equation}
can consistently account for the occurrence of SSB in quantum field
theory. However, at any finite value of the ultraviolet cutoff 
$\Lambda\sim 1/a$ there is no reason why the mass gap should vanish
{\em before} the system being in the broken phase,  thus opening the
possibility that $r_s>r_c$. As discussed in Ref.~\cite{zeit,zappa}, by
exploring the $m$-dependence of the effective potential of the cutoff
theory in those approximations consistent with ``triviality'', this
indeed occurs and one finds a first order phase transition so that the
massless regime lies in the broken phase (see
Eqs.(\ref{eq:1},\ref{eq:4})). For convenience of the reader, in the
case of the Gaussian approximation,
we describe in the Appendix the evaluation of the effective potential
and its dependence on the bare mass parameter.

At the same time\cite{rit,zeit,zappa}~,
SSB requires $m$, the mass gap in the symmetric phase, to vanish in
units of $m_h$, the mass gap of the broken phase, in the continuum
limit $t\to \infty$  consistently with Eq.(\ref{eq:36}) 
(see Eq.~(\ref{eq:a13}) of the Appendix).

As anticipated in the Introduction,  a definitive test of the validity
of this theoretical framework  requires to explore the shape of the
effective potential after an {\em a priori} determination of $r_c$. To
this end we shall use the analysis of lattice data presented by Brahm
in Ref.~\cite{brahm}. By considering data from different groups and
different lattices,  including a parametrization of the finite-size
effects and  extrapolating the lattice data down to zero mass, Brahm's
analysis provides a rather precise determination of the value of $r_c$
at $\lambda_0=0.5$:
\begin{equation}
\label{eq:37}
r_c=-0.2240 -{{1.00\pm 0.05}\over{L^2}} \pm 0.0010  \,.
\end{equation}
For L=16, Eq.~(\ref{eq:37}) predicts
\begin{equation}
\label{eq:38}
r_c=-0.2279 (10) \,.
\end{equation}
In order to check this prediction we have performed a direct
numerical calculation of $r_c$ at $\lambda_0=0.5$
on a $16^4$ lattice.

An accurate procedure to determine $r_c(\lambda_0)$ uses the
susceptibility $\chi$:
\begin{displaymath}
\chi=L^4 \left[ \left\langle |\Phi|^2 \right\rangle - 
\left\langle |\Phi| \right\rangle^2 \right] \,
\end{displaymath}
where
\begin{displaymath}
|\Phi| = \frac{1}{L^4} \sum_x | \Phi(x) | \,.
\end{displaymath}
Indeed it is known that~\cite{luscher87} near the critical region 
$\chi^{-1} \sim (r_c-r_0)$, modulo logarithmic corrections. However,
the logarithmic modifications of the free field scaling law are
important only when $r_0$ is very close to $r_c$. Accordingly, we have
measured the susceptibility in the symmetric phase $r_0 < r_c$ and
fitted linearly the inverse of the susceptibility. We find that the
linear fit is rather good for $0.24 \le r_0 \le 0.35$ (see Fig.~1).
As a result we obtain
\begin{equation}
\label{eq:rcritico}
r_c =  -0.22788 (138) \,.
\end{equation}
%
%
%
\begin{figure}[t]
\epsfig{file=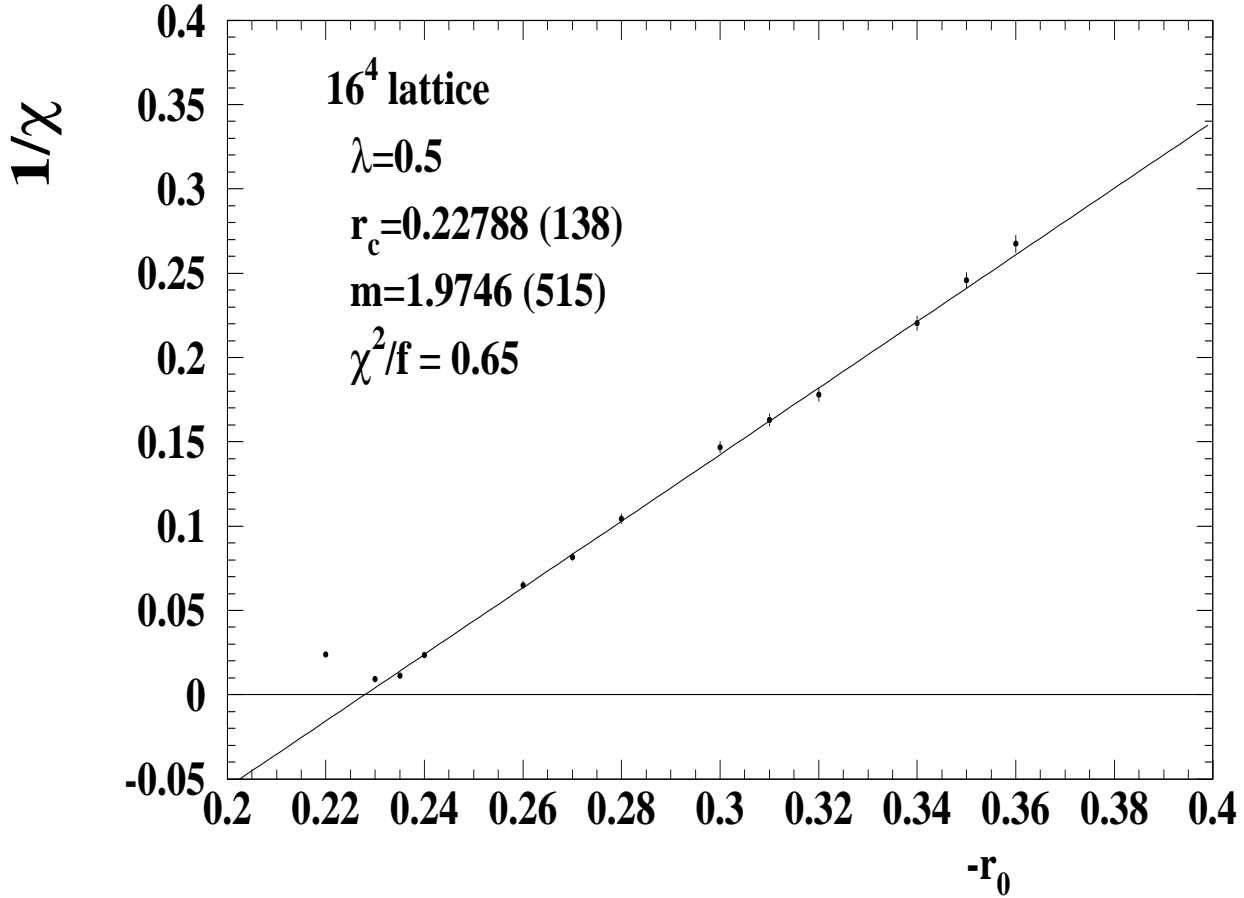,width=\textwidth,height=12truecm}
\caption{The inverse of the susceptibility versus $-r_0$.
The solid line is the  linear fit  $m(r_c-r_0)$.}
\label{fig:fitrc}
\end{figure}
%
%
%
The agreement
with Brahm's prediction is remarkable
and, thus, we have
two independent indications that the
central value $r_c=-0.2279$ represents the input definition
of the massless regime for a numerical computation of the slope of the
effective potential on a $16^4$ lattice with the action
Eq.~(\ref{eq:25}) at $\lambda_0=0.5$.  
Finally, after obtaining
$J=J(\phi_B)$ in a  model-independent way, we shall compare the
lattice data with the two alternative descriptions of the phase
transition in $(\lambda\Phi^4)_4$.

\begin{figure}[t]
\epsfig{file=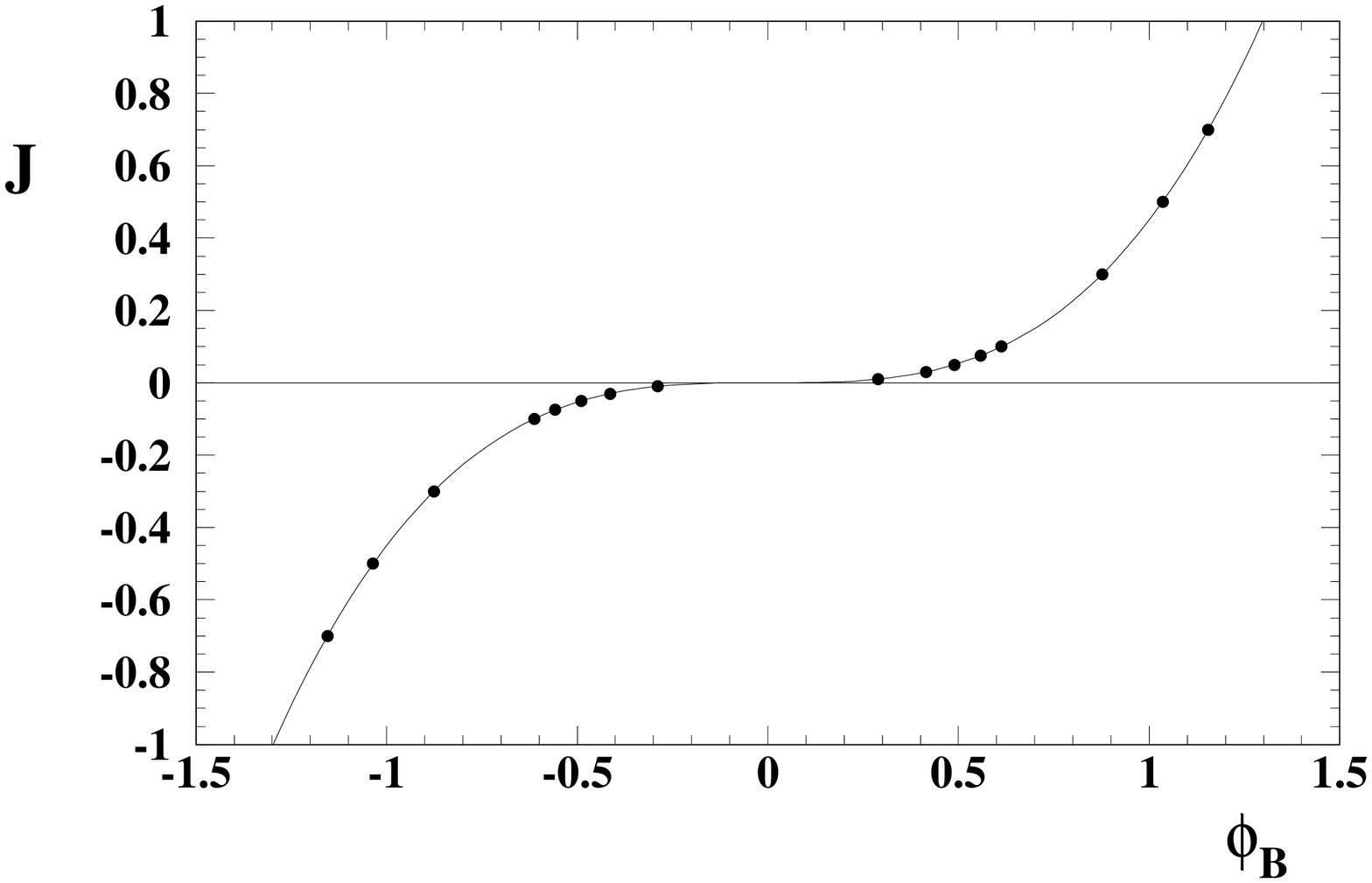,width=\textwidth,height=12truecm}
\caption{The external current $J$ versus $\phi_B$. The solid
line is Eq.~(\protect\ref{eq:39}) with $\beta=\delta=0$ and
$\alpha=1.520 \times 10^{-2}$, $\gamma=0.44960$.}
\label{fig:fit}
\end{figure}

For our Monte Carlo simulation we used the standard Metropolis
algorithm to update the lattice configurations weighted by the action
Eq.~(\ref{eq:25}). In order to avoid the trapping into metastable
states due to the underlying Ising dynamics we followed the upgrade of
the scalar field $\Phi(x)$ with the upgrade of the sign of $\Phi(x)$.
This is done according to the effective Ising action~\cite{Brower89}
%
\begin{displaymath}
S_{\mathrm{Ising}}  =   J \, \sum_x \left| \Phi(x) \right| s(x) 
 \mbox{ } -  \sum_x \sum_{\hat{\mu}}
\left| \Phi(x+\hat{\mu}) \Phi(x) \right|
s(x+\hat{\mu}) s(x) \;,
\end{displaymath}
where $s(x) = \mathrm{sign}(\Phi(x))$. We measured the vacuum
expectation  value of the scalar field 
\begin{displaymath}
\langle\Phi\rangle_J={{1}\over{N_c}}\sum^{N_c}_{i=1}{{1}\over{L^4}}
\sum_x\Phi^i(x)
\end{displaymath}
where $N_c$ is the number of configurations generated with the
action~(\ref{eq:25}),  for 16 different values of the external source
in the range $0.01\leq |J| \leq 0.70$. Statistical errors are
evaluated using the jackknife algorithm~\cite{Efron82}  adapted to
take into account the correlations between consecutive lattice
configurations \cite{Berg89}. Our final results for
$\langle\Phi\rangle_J=\phi_B(J)$ are shown in  Table~\ref{table:I}.
\begin{table}
\tabcolsep .2cm
\renewcommand{\arraystretch}{2}
\begin{center}
\begin{tabular}{|c|l||c|l|}
\hline
{$J$}       &    {$\phi_B(J)$}   &  {$J$}  &   {$\phi_B(J)$}  \\ \hline
-0.010      &   -0.288862 (695)  &  0.010  &   0.289389 (787) \\ \hline
-0.030      &   -0.413565 (321)  &  0.030  &   0.414713 (376) \\ \hline
-0.050      &   -0.488797 (296)  &  0.050  &   0.489132 (249) \\ \hline
-0.075      &   -0.557737 (181)  &  0.075  &   0.557961 (182) \\ \hline
-0.100      &   -0.612497 (169)  &  0.100  &   0.612865 (151) \\ \hline
-0.300      &   -0.876352 (111)  &  0.300  &   0.876518 (95)  \\ \hline
-0.500      &   -1.03526~ (8)    &  0.500  &   1.03532~ (7)   \\ \hline
-0.700      &   -1.15518~ (8)    &  0.700  &   1.15528~ (7)   \\ \hline
\end{tabular}
\caption{We report the values of $\phi_B(J)$  as obtained with our
$16^4$ lattice at $\lambda_0=0.5$ and $r_0=r_c=-0.2279$. Errors are
statistical only.}
\end{center}
\label{table:I}
\end{table}
We have compared these data with the two alternative approaches
motivated by ``triviality'' or  based on perturbation theory. In the
former case we have chosen the general form
\begin{equation}
\label{eq:39}
J(\phi_B)=\alpha\phi^3_B\ln(\phi^2_B)+\beta\phi_B+\gamma\phi^3_B
+\delta\phi^3_B\ln^2(\phi^2_B)
\end{equation}
which, besides including an explicit scale breaking term $\beta$
accounts for possible deviations from the ``triviality'' structure in
Eqs.(\ref{eq:1},\ref{eq:4}). Indeed, the presence of residual, genuine
self-interaction effects for the field $h(x)$ may show up in a non
vanishing coefficient $\delta$ of the $\ln^2(\phi^2_B)$ term as
predicted, for instance, from the structure of the effective potential
in a perturbative two-loop calculation. Finally, since we expect to be
very close to the massless regime, the effect of a non vanishing mass
scale $\pm M^2$ in the argument of the logarithms (apart from those
effectively  included in the coefficient $\beta$) should be
negligible. A consistency check of this assumption can be obtained,
apart from the quality of the fit, from the size of the coefficient
$\beta$ since, away from criticality, one has  $\beta\sim\pm M^2$. The
results of the fit to the data in Table I with  Eq.~(\ref{eq:39}) is
\begin{eqnarray*}
\alpha  & = & (1.535\pm 0.062)\cdot 10^{-2}             \\
\beta   & = & (0.3\pm 6.3)\cdot 10^{-4}                 \\
\gamma  & = & 0.44955\pm0.00061                         \\
\delta  & = & (1.3\pm 6.4)\cdot 10^{-4}                 \\
\frac{\chi^2}{{\mathrm{d.o.f}}} & = & \frac{14.4}{16-4}
\end{eqnarray*}
It is clear that the model-independent calculation of the slope of the
effective potential, is in very good agreement with our
predictions based on Eqs.~(\ref{eq:1},\ref{eq:4}). In fact, the fit
shows no evidence  for non-vanishing coefficients $\beta$ and $\delta$.
By constraining  $\beta=\delta=0$ in the fit, we obtain 
$\alpha=1.520(18)\times 10^{-2}$, $\gamma=0.44960(7)$,
$\chi^2/{\mathrm{d.o.f}}=15.0/(16-2)$.
In Figure~\ref{fig:fit} we display our data together with the best fit
Eq.~(\ref{eq:39}), with the constraint $\beta=\delta=0$.
%
%
Also, the size of the coefficient
$\beta$ is so small that the effect of  a non vanishing $\pm M^2\sim
\beta$ in the arguments of the logarithms is  totally negligible, at
least in the explored range of $\phi_B$, $\phi^2_B>0.08$. Notice that
the only finite size effect included in our  analysis is the $1/L^2$
correction to the Brahm's value for the critical bare mass. Therefore,
the excellent $\chi^2$'s of the fits show that, at least for $|J|\ge
0.01$, our $16^4$ lattice behaves as an infinite system.

Let us now compare the lattice data in Table~\ref{table:I} with a
perturbative evaluation of the effective potential. Indeed, the small
size of the bare coupling  $\lambda_0/\pi^2\sim 0.05$ might suggest
that the agreement between the lattice data and
Eqs.~(\ref{eq:1},\ref{eq:4}) represents just a trivial test of
perturbation theory. To this end we have compared with the full
two-loop calculation of  Ref.~\cite{jones}. In the dimensional
regularization scheme, their expression for the  effective potential
in the massless regime and for the one component theory is
($m^2_2=\lambda\phi^2_B/2$, 
$\Omega(1)={{3}\over{4}}S-{{1}\over{3}}\zeta(2)$ with
$S=1/2^2+1/5^2+1/8^2+\ldots$ and $\overline {\ln}$ includes in the
definition of the logarithm additional  terms of the MS scheme)
\begin{equation}
\label{eq:40}
V^{\mathrm{2-loop}}(\phi_B)=V_0(\phi_B)+V_1(\phi_B)+V_2(\phi_B)
\end{equation}
with
\begin{eqnarray*}
V_0(\phi_B) & = & \frac{\lambda}{4!}\phi^4_B       \\
V_1(\phi_B) & = &
\frac{1}{64\pi^2}m^4_2[\overline{\ln}\frac{m^2_2}{\mu^2}-\frac{3}{2}]
\end{eqnarray*}
and
%
\begin{displaymath}
V_2(\phi_B)  =   \frac{1}{256\pi^4} \frac{\lambda^2\phi^2_Bm^2_2}{8}       
\left[ 
5 + 8\Omega(1)-4\overline{\ln}\frac{m^2_2}{\mu^2}  
+ \overline{\ln}^2\frac{m^2_2}{\mu^2} \right]
+  \frac{1}{256\pi^4} \frac{\lambda m^4_2}{8}
\left[ 1-\overline{\ln}\frac{m^2_2}{\mu^2} \right]^2  
\end{displaymath}
By transforming to our notations, namely
\begin{eqnarray*}
\lambda  & \equiv &  6\lambda_0  \\
\overline{\ln}(m^2_2/\mu^2)-{{3}\over{2}} & \equiv &
\ln(m^2_2/\Lambda^2)-{{1}\over{2}}
\end{eqnarray*}
and computing $J^{\mathrm{2-loop}}(\phi_B)=dV^{\mathrm{2-loop}}/d\phi_B$
we find the final expression
%
\begin{equation}
\label{eq:41}
J^{\mathrm{2-loop}}(\phi_B)=
\frac{\lambda_0\phi^3_B}{1+\frac{9\lambda_0}{16\pi^2}
\ln \frac{\Lambda^2}{3\lambda_0\phi^2_B} } 
+\frac{\lambda_0^3\phi^3_B}{256\pi^4}
\left[
27\ln \frac{\Lambda^2}{3\lambda_0\phi^2_B} + 54 + 432\Omega(1) \right]
\end{equation}
in which we have used the perturbative $\beta$ function to resum the
leading logarithmic terms to all orders. By fitting the lattice data
in Table~\ref{table:I}  with Eq.~(\ref{eq:41}) for $\lambda_0=0.5$,
and leaving out the scale $\Lambda$ as a free parameter in the fit, we
find the result for the 2-loop, leading-log improved fit
\begin{displaymath}
\left( \frac{\chi^2}{\mathrm{d.o.f}} \right)_{\mathrm{2-loop}} =
\frac{1162}{16-1}
\end{displaymath}
which, indeed, shows that, despite of the small value of the bare
coupling, perturbation theory (in one of its more refined versions) is
totally unable to describe the lattice data. As anticipated in the
Introduction, this result is not unexpected  and, therefore, the
agreement with  Eqs.(\ref{eq:1},\ref{eq:4}) is not a trivial test of
perturbation theory but provides, rather, a non-perturbative test of
``triviality''. The crux of the matter has to be found in the {\em
qualitative} conflict between Eq.~(\ref{eq:39})  (for
$\beta=\delta=0$) and Eq.~(\ref{eq:41}). In the former case,  based on
a first-order description of the phase transition, the massless 
regime lies in the broken phase and there are non-trivial minima for
the  effective potential so that $J$ vanishes at non zero values
$\phi_B=\pm v_B$. On the other hand, Eq.~(\ref{eq:41}), consistent 
with a second-order phase transition,  can only  vanish at $\phi_B=0$
since, within the perturbative approach, one needs a non-vanishing and
negative renormalized mass squared $-M^2$ to obtain SSB. Our lattice
data for the response of the system to the external source are precise
enough to detect the sizeable difference produced by the two
extrapolations towards $J=0$. 

Finally, to perform an additional check, we have used the complete
form for $V^{\mathrm{2-loop}}(\phi_B)$ reported in Eq.~(\ref{eq:40}) by
allowing for the presence of a negative mass parameter $-M^2$ as in
Ref.~\cite{jones}. In this case, where the classical potential becomes
\begin{displaymath}
V_0(\phi_B)={{\lambda}\over{4!}}\phi^4_B-{{1}\over{2}}M^2\phi^2_B
\end{displaymath}
and the mass parameter $m^2_2$ has to be replaced everywhere by
\begin{displaymath}
m^2_2={{\lambda\phi^2_B}\over{2}}-M^2 \,,
\end{displaymath}
we still obtain an extremely poor fit
\begin{displaymath}
\left( \frac{\chi^2}{\mathrm{d.o.f}} \right)_{\mathrm{2-loop}}=
\frac{152}{16-2} \,,
\end{displaymath}

Summarizing: our analysis predicts that, for
$\lambda_0=0.5$, the massless regime of $(\lambda\Phi^4)_4$ 
corresponds to $r_0=r_c=-0.2279$ on a $16^4$ lattice. The resulting
effective potential, computed in a fully model-independent way, is in
excellent agreement with the general ``triviality'' structure in 
Eqs.~(\ref{eq:1},\ref{eq:4}) and cannot be reproduced  in a
perturbative expansion, despite of the small value of the bare
coupling $\lambda_0/\pi^2 \sim 0.05$. Our analysis, while
confirming the existence on the lattice of a remarkable phase of
$(\lambda\Phi^4)_4$ where SSB is generated through ``dimensional
transmutation'' \cite{CW}, enforces the first numerical  evidences of
Ref.~\cite{andro} pointing out the inner contradiction between 
perturbation theory and ``triviality''. The most important
consequences of this basic inadequacy will be illustrated in detail in
Sect.~3

\section{Conclusions and outlook}

Let us now compare our results with the output of the existing lattice
simulations. So far,  the theoretical expectations based on
``triviality'' have been numerically confirmed and there is
overwhelming evidence that all {\em observable} interaction effects
vanish when approaching the continuum limit, i.e. when the  physical
correlation length of the broken phase becomes large in units of the
lattice spacing. 

In all interpretations of the lattice simulations performed so far the
validity of the perturbative relation~(\ref{eq:23}) has been assumed. 
Let us refer to the very complete review by C.B.Lang \cite{lang}.
There, for the O(4) theory, the value of $Z=Z_h$ is extracted from the
large distance decay of the Goldstone boson propagator and used in
Eq.~(\ref{eq:23}) to define a renormalized VEV $u_R$ from the average
bare VEV $v_B$  measured on the lattice.  Quite independently of any
interpretation, the lattice data provide $Z_h=1$ to very good accuracy
(see Tab.II in Ref.~\cite{lang}) and completely confirm the trend in 
Eqs.~(\ref{eq:7},\ref{eq:22}) (see Fig.19 in Ref.~\cite{lang}) so
that, in this approach, one finds
\begin{equation}
\label{eq:42}
\frac{m^2_h}{u^2_R} \sim \frac{m^2_h}{v^2_B} \sim 
\frac{8\pi^2}{3 \ln(\Lambda/m_h) } \to 0  \,.
\end{equation}
The above relation has been interpreted so far on the basis of the
leading-log formula for the running coupling constant. Indeed, in
perturbation theory, where one relates $u_R\sim v_B$ to $m_h$ through
the renormalized 4-point function at external momenta comparable to
the Higgs mass itself, 
\begin{equation}
\label{eq:43}
{{m^2_h}\over{u^2_R}}\sim 3 \lambda_R(m^2_h)  \,, 
\end{equation}
in the leading-log approximation
\begin{equation}
\label{eq:44}
\lambda_R(m^2_h) =
\frac{\lambda_0}{1+
\frac{9\lambda_0}{8\pi^2} \ln\frac{\Lambda}{m_h} }
\end{equation}
one deduces Eq.~(\ref{eq:42}) for $\lambda_0\to \infty$. Therefore,
when $u^2_R \equiv v^2_B/Z_h \sim v^2_B$ is kept fixed as a
cutoff-independent quantity (and  related to the Fermi constant
through $u^2_R\sim 1/G_F\sqrt{2}$),  and the result~(\ref{eq:22}) is
interpreted within perturbation theory, one concludes that SSB is only
possible in the presence of an  ultraviolet cutoff (this point of view
leads to upper bounds on the  Higgs mass \cite{call,lang,neu} $m_h
<700-900$ GeV). 

As pointed out in the  Introduction, however, this interpretation of
``triviality'' is, at least, suspicious. Indeed, within perturbation
theory itself, it is in contradiction with explicit two-loop
calculations of  $\beta_{\mathrm{pert}}$. In this case, from
Eq.~(\ref{eq:8}) one would deduce that the bare coupling flows toward
the ultraviolet fixed point
\begin{equation}
\label{eq:45}
\lim_{\Lambda\to \infty}~{{\lambda_0(\Lambda)}\over{\pi^2}}
={{24}\over{17}}={{\lambda^*}\over{\pi^2}}
\end{equation}
for {\em any} $0<\lambda_R<\lambda^*$  and, by assuming the validity
of Eq.~(\ref{eq:43}), there is no reason why $\lambda_R$ and $m_h$
should vanish in the limit $\Lambda\to \infty$ ($\lambda^*$, in any
case, disappears at 3-loop but reappears at 4-loops \cite{beta}).

Quite independently of this remark, which rather concerns the internal 
consistency of the perturbative approach to ``triviality'',  our
numerical results of Sect.~2 (and those of Ref.~\cite{andro}), show
that in the response of the  lattice theory to the external source $J$
there is no trace of any residual  $h$-field self-interaction effect
but the lattice effective potential cannot be reproduced in
perturbation theory. Therefore,  {\em any} perturbative interpretation
of ``triviality'' can hardly be taken as correct.  Close to the
continuum limit as one can be, we do find a  definite numerical
evidence for non trivial minima of the effective potential, and, as
such, there is no reason why SSB should not coexist with
``triviality''. However, the Higgs mass $m_h$ defined from the vacuum
energy in Eq.~(\ref{eq:3}), which determines the physical scale of the
broken phase,  is quite unrelated to $\lambda_R$, which vanishes, and
does not represent a measure of any interaction. 

Before exploring the consequences of our picture of SSB, however, let
us briefly discuss the case of a $(\lambda\Phi^4)_4$ theory in the
continuous symmetry O(N) case. The use of radial and angular fields
allows to deduce easily the structure of the effective potential for
the O(N) theory. In fact, as discussed in
Ref.~\cite{con,new,resolution}, one expects
Eqs.~(\ref{eq:20},\ref{eq:21}) of the  one-component theory, to be
also valid for the radial field in the O(N)-symmetric case. The
explanation for this result is extremely intuitive and originates from
Ref.~\cite{dj} which obtained, for the radial field, the same
effective potential as in the discrete-symmetry case. The
Goldstone-boson fields contribute to the effective potential only
through their zero-point energy, that is an additional constant,
since, according to ``triviality'', they are  free massless fields.
Thus, in the O(2)-symmetric case, one may take the diagram
$(V_{\mathrm{eff}},\phi_B)$ for the one-component theory and ``rotate'' it
around the $V_{\mathrm{eff}}$ symmetry axis. This generates a 
three-dimensional diagram $(V_{\mathrm{eff}},\phi_1,\phi_2)$ 
where $V_{\mathrm{eff}}$,
owing to the O(2) symmetry, only depends on the bare radial field,
\begin{equation}
\label{eq:46}
\rho_B=\sqrt{\phi^2_1+\phi^2_2}
\end{equation}
in exactly the same way as $V_{\mathrm{eff}}$ depends on $\phi_B$ in the
one-component theory; namely
($\omega^2(\rho^2_B)=3\tilde{\lambda}\rho^2_B$)
\begin{equation}
\label{eq:47}
V_{\mathrm{eff}}(\rho_B)  =  \frac{\tilde{\lambda}}{4} \rho^4_B + 
\frac{\omega^4(\rho^2_B)}{64\pi^2}  \left( \ln \frac{\omega^2
(\rho^2_B) }{\Lambda^2} -  \frac{1}{2} \right) \,. 
\end{equation}
This has been explicitly checked in Ref.~\cite{andro} with the Monte
Carlo  simulation of the O(2) lattice theory employing the action
%
\begin{equation}
\label{eq:48}
S = \sum_x \sum^2_{i=1} 
\left[ 
\frac{1}{2} 
\sum_{\hat{\mu}}
\left( \Phi_i(x+\hat{\mu}) -  \Phi_i(x) \right)^2 
+ \frac{1}{2}r_0
\left( \Phi_i(x)\Phi_i(x) \right) 
+ \frac{\lambda_0}{4} \left( \Phi_i(x)\Phi_i(x) \right)^2 -
J_i\Phi_i(x) \right]
\end{equation}
where $\Phi_1$ and $\Phi_2$ are coupled to two constant external
sources  $J_1$ and $J_2$. By using $J_1=J\cos\theta$ and
$J_2=J\sin\theta$ and having defined
$\phi_{1}=\langle\Phi_{1}\rangle_{J_1,J_2}$,
$\phi_{2}=\langle\Phi_{2}\rangle_{J_1,J_2}$, 
one can compute the bare radial field Eq.~(\ref{eq:46})
\begin{equation}
\label{eq:49}
\rho_B=\rho_B(J)
\end{equation}
and invert Eq.~(\ref{eq:49}) to obtain the slope of the effective
potential. As shown in Ref.~\cite{andro} the lattice data for
$\rho_B=\rho_B(J)$ in the  massless regime are remarkably reproduced
by
\begin{equation}
\label{eq:50}
J(\rho_B)={{dV_{\mathrm{eff}}(\rho_B)}\over{d\rho_B}}=
{{9\tilde{\lambda}^2\rho^3_B}\over{16\pi^2
}}\ln{{\rho^2_B}\over{v^2_B}} \,,
\end{equation}
thus providing definite numerical support for the exactness conjecture
of Eqs.(\ref{eq:20}, \ref{eq:21}) \cite{resolution,zeit}.

Finally, in the post Gaussian calculation of Ref.~\cite{rit2},  the
numerical solution of the integral equation for the shifted radial
field propagator gives for the Higgs mass $m_h=$2.21 TeV for N=2 and
$m_h=$2.27 TeV for N=4 to compare with the prediction of
Eq.~(\ref{eq:21}) $m_h=$2.19 TeV for  $v_R\sim $246 GeV if the
physical VEV introduced in Sect~.2 is related to the Fermi constant
$G_F$ in the usual way.

Therefore, in the most appealing theoretical framework, where SSB is
generated from a theory which does not possess any physical scale in
its symmetric phase $\langle\Phi\rangle=0$, we end up with a definite
prediction for the Higgs mass, namely $m_h\sim$2.2 TeV
\cite{con,iban,new,resolution,zeit}. In general, the Higgs can be
lighter or heavier \cite{zeit}, depending on the flow of the bare mass
$r_0=r_0(t)$ in the continuum limit $t\to\infty$. In this case one
ends up with the more general result \cite{zeit}
\begin{equation}
\label{eq:51}
m^2_h=8\pi^2\zeta v^2_R
\end{equation}
with
\begin{equation}
\label{eq:52}
0<\zeta\leq 2 \;.
\end{equation}
$\zeta=1$ corresponds to $r_0(t)=r_c(t)$ and $\zeta=2$ to 
$r_0(t)=r_s(t)>r_c(t)$, as defined in Sect.~2 (see also the Appendix).
This limiting situation, namely
when $m_h/v_R=4\pi $ and symmetric and broken phases
have the same energy, places an upper bound on the Higgs mass $m_h<$3.1 TeV.
 The range $0<\zeta<1$,
corresponding to values $r_0(t)<r_c(t)$, for which the theory cannot
be quantized in its symmetric phase $\langle\Phi\rangle=0$, is allowed
by the  Renormalization Group analysis of the effective potential
\cite{zeit} and cannot be discarded.

The existence of an upper limit for the ratio $m_h/v_R$ from vacuum
stability (and {\em not} from `triviality')
is a genuine quantum phenomenon which has no counterpart in the
semiclassical `double well' picture. It is a direct consequence of
the first-order phase transition and is in qualitative agreement with
similar extimates based on the Bethe-Salpeter equation.
By investigating the condition for a zero-mass bound state
in Higgs-Higgs scattering, signaling
the instability of the spontaneously broken phase,
one finds the result $m_h\lesssim 2.4$ TeV \cite{rupp} or
$m_h\lesssim 3.4$ TeV
\cite{dilcher} in different approximations of the Bethe-Salpeter
kernel.

Obviously, the above extimates are only valid provided the
contribution to the effective potential from the other fields, namely
gauge bosons and fermions, is negligible and can be treated as a small
perturbation, as in the original formulation of the Weinberg-Salam
theory \cite{WS} where SSB is generated in the pure  scalar sector
(this is certainly possible for $\zeta\sim 1$ where the Higgs mass is
large compared to the top quark mass as measured by  the CDF and D0
Collaborations, $m_t=180\pm12$ GeV \cite{cdf}).

However, even though the Higgs mass would turn out to be considerably
lighter than our reference value 2.2 TeV, there are substantial
implications. In fact, the Higgs phenomenology, on the basis of gauge
invariance, depends on the details of the pure scalar sector. For
instance, consider the Higgs decay width to $W$ and $Z$  bosons. The
conventional calculation would give a huge width, of  order $G_F m_h^3
\sim m_h$ for $m_h\sim$ 1 TeV. However, in a renormalizable-gauge 
calculation of the imaginary part of the Higgs self-energy, this
result comes from a diagram in which the Higgs supposedly  couples
strongly to a loop of Goldstone bosons with a {\em physical} strength
proportional to its mass squared, an effect which, in principle, has
nothing to do with the  gauge sector but crucially depends on the
description of the pure scalar theory at zero gauge coupling. Now, if
``triviality'' is true, as we believe, {\em all} interaction effects
of the pure $\lambda\Phi^4$ sector of the standard model have to be
reabsorbed into two numbers, namely $m_h$ and $v_R$, and there are no
residual interactions. However,  if a Higgs particle can decay into
two Goldstone bosons {\em there are} observable interactions, namely
there is a non-trivial scattering matrix for Goldstone-Goldstone
scattering with a pole in the  complex plane whose real and imaginary
parts are related to the Higgs mass and to the Higgs decay width. 
Beyond perturbation theory, this process cannot be there and,
therefore, if ``triviality'' is true, a heavy Higgs {\em must be} a
relatively  narrow resonance, decaying predominantly to $t \bar{t}$
quarks.  

In conclusion, according to our analysis ``triviality'' implies just
the opposite of what is generally believed,  namely SSB with
elementary scalar fields, and as such the essential ingredient for the
Higgs mechanism, poses, by itself, no problems of internal consistency
as far as its quantum field theoretical limit is concerned. Truly
enough, this is only relevant to the continuum theory and, therefore,
we have little to say if the scalar sector of the standard model 
turns out to be a low-energy effective description of symmetry
breaking. In this case, a perturbative approach in terms of a light
Higgs mass (say $m_h\lesssim 200$ GeV \cite{cabibbo}) is still
acceptable, since the scale at which the picture  breaks down is
exponentially decoupled from the Higgs mass, and physically equivalent
to our description in the limit $\zeta<< 1$.  On the other hand, if
the Higgs turns out to be very heavy a measure of its  decay width
will represent a test of our predictions and the experiment will 
decide between the two descriptions.\\
\vspace{0.5cm}
\begin{center}
{\em Acknowledgement}
\end{center}
We thank A. Agodi, G. Andronico, P.M. Stevenson, and D. Zappal\`a for
many useful discussions.

\section*{Appendix}

 Let us now address the question of the stability of the symmetric 
phase in the Gaussian approximation. This is a particularly simple type
of calculation and has the advantage, in a `trivial' theory such as
$(\lambda\Phi^4)_4$, of being effectively exact, in its renormalized
form, as discussed in Sect.~1. 

 The starting point for our analysis is the effective
potential for composite operators
introduced by Cornwall, Jackiw and Tomboulis (CJT)~\cite{cjt} which 
represents a powerful analytic
tool to investigate the structure of the effective potential beyond perturbation
theory.

The CJT approach 
is based on the exact
relation between the effective potential and the energy functional
for a constant bare field configuration and arbitrary equal-time propagator
$G(\vec x, \vec y)=G(x,y)\big |_{x_0=y_0}$
\begin{displaymath}
     \int d^3x~V_{\rm eff}(\phi_B)=E[\phi_B,G_0(\phi_B)]
\end{displaymath}
where
\begin{displaymath}
             E[\phi_B,G]={\rm min}_{\Psi}~\langle\Psi|H|\Psi\rangle
\end{displaymath}
with the conditions
\begin{displaymath}
                       \langle\Psi|\Psi\rangle=1
\end{displaymath}
\begin{displaymath}
                 \langle\Psi|\Phi|\Psi\rangle=\phi_B
\end{displaymath}
\begin{displaymath}
   \langle\Psi|\Phi(\vec x)\Phi(\vec y)|\Psi\rangle=\phi^2_B+ G(\vec x,\vec y)
\end{displaymath}
and
\begin{equation}
\label{eq:a1}
\left. 
\frac{\delta E[\phi_B,G]}{\delta G(\vec x,\vec y)} \right|_{G=G_0(\phi_B)}=0
\,.
\end{equation}
Finally, the absolute minima of $V_{\rm eff}$, say $\pm v_B$, define the ground
state energy density $W$ up to an arbitrary additive constant (for which we 
shall
take the value at $\phi_B=0$) so that
\begin{displaymath}
         W=V_{\rm eff}(\pm v_B)-V_{\rm eff} (0)
\end{displaymath}
and SSB corresponds to the situation $W<0$ for $v_B\neq 0$.

In general one may employ different approximation schemes
to $E[\phi_B,G]$ and ,
therefore, to $V_{\rm eff}$. For instance, one may adopt the modified loop
expansion discussed in~\cite{cjt} in terms of two-particle irreducible 
vacuum-vacuum graphs with vertices determined by the shifted interaction
Lagrangian and propagators fixed by $G(x,y)$. This type of `loop-expansion',
however,
is very different from the usual loop-expansion in powers of $\hbar$
for the effective
potential since, for instance, the inclusion of
a single graph in $E[\phi_B,G]$ produces, through the 
(in general non perturbative) solution of Eq.~(\ref{eq:a1}), an infinite number of
graphs in terms of the 1-loop propagator
\begin{equation}
\label{eq:a2}
D(x,y;\phi_B)=\int {{d^4p}\over{(2\pi)^4}}{{1}\over{p^2+r_0+3\lambda_0\phi^2_B}}
\exp[ip(x-y)] .
\end{equation}
We want to emphasize that a
systematic expansion in powers of $\hbar$ is precisely equivalent to
a weak coupling expansion in the shifted theory and, as such, unable to 
provide any meaningful indication for a theory
where no {\it observable} interaction effect can survive in the continuum 
limit. In fact, the attempt to renormalize $(\lambda\Phi^4)_4$ in the 
standard
perturbative approach is based on the concept of a cutoff independent and
non-vanishing renormalized coupling at non zero external momenta
$\lambda_R(Q^2)$, a concept for which there is no room in a `trivial' theory
(in order to fully realize the essentially perturbative nature of the usual loop
expansion we address the interested reader to ref.\cite{new} where the
meaning of the various contributions and their relation with the perturbative
expansion are clearly illustrated). 

It is clear from its definition that $G(x,y)$ is the propagator of the
shifted field $h(x)$ for each given value of the vacuum field $\phi_B$. 
Thus, a consistency requirement for SSB
in $(\lambda\Phi^4)_4$ is that, in the continuum limit, $G(x,y)$ 
has to reduce to a free field propagator in agreement with the basic 
`triviality' results of Sect.~2.
Now, 
it is well known that in perturbation theory this does not occur. Beyond the
lowest order 1-loop approximation in Eq.~(\ref{eq:a2}), 
the shifted field propagator has ultraviolet divergent
corrections implying a non trivial cutoff-dependence of $Z_h$ for
the quantum field $h(x)$ and as such a non-trivial anomalous dimension
\begin{displaymath}
\gamma_h(\lambda_0)=-{{1}\over{2}}{{\partial\ln Z_h}\over{\partial\ln \Lambda}}=
{{3\lambda^2_0}\over{256\pi^4}}+O(\lambda^3_0)+ \cdots 
\end{displaymath}
 which cannot vanish in the continuum limit $\Lambda \to \infty$
 unless, in the same limit,
 $\lambda_0=\lambda_0(\Lambda)\to 0$, in contrast with the
results obtained from the perturbative $\beta$-function.

On the basis of the previous discussion, it should be clear that `triviality'
forces to define the continuum theory starting from a regularized
version where all shifted field self-interaction 
effects are neglected 
or become {\it unobservable}, being reabsorbed into its mass.
 Therefore, only
approximations to $E[\phi_B,G]$ fulfilling the consistency requirement 
\begin{displaymath}
        G(x,y)\to
\int {{d^4p}\over{(2\pi)^4}}{{1}\over{p^2+m^2_h}}
\exp[ip(x-y)]
\end{displaymath}
when the ultraviolet regulator is removed, are allowed. A particularly simple
class of approximations is provided by the 1-loop and Gaussian effective 
potential where, by definition, the shifted field $h(x)$ is governed by a
quadratic Hamiltonian consistently with ``triviality''.
 
The Gaussian effective potential has been considered as a convenient method to
study vacuum stability beyond perturbation theory by many authors 
and is based on the choice of the Gaussian 
wave-functionals
%
\begin{displaymath}
\Psi_G[\Phi]=  ({\rm Det}~G)^{-1/4} \,
\times  \, \exp \left[ -\frac{1}{4} \int d^3x\int d^3y \left( 
\Phi(\vec x)-\phi_B \right) G^{-1}(\vec x, \vec y)
\left( \Phi(\vec y)-\phi_B \right) \right] 
\end{displaymath}
where
\begin{displaymath}
G(\vec x, \vec y)=\int {{d^3p}\over{(2\pi)^3}}~ 
{{\exp[i\vec p\cdot(\vec x-\vec y)]}\over{ 2\sqrt{{\vec p}^2+\omega^2} } }
\end{displaymath}
$\omega$ being a variational parameter.
In this class of states one obtains \cite{cjt,stevenson,cian}
\begin{displaymath}
E_G(\phi_B,\omega)=\int d^3x~ V_G(\phi_B,\omega)
\end{displaymath}
with
%
\begin{equation}
\label{eq:a3}
V_G(\phi_B,\omega)=\frac{1}{2} r_0 \phi^2_B + \frac{1}{4} \lambda_0 \phi^4_B +
I(x,x)
+
\frac{1}{2} \left( r_0 - \omega^2 + 3 \lambda_0 \phi^2_B \right) G(x,x) +
\frac{3 \lambda_0}{4} G(x,x) G(x,x)
\end{equation}
where
\begin{equation}
\label{eq:a4}
I(x,y)=\frac{1}{2} G^{-1}(x,y)  \,.
\end{equation}
In Eqs.~(\ref{eq:a3},\ref{eq:a4}) we have used the 4-dimensional
notation  to express $G(x,x)$ and $I(x,x)$ in terms of the
4-dimensional Euclidean cutoff (appropriate for the lattice theory)
as
\begin{equation}
\label{eq:a5}
G(x,x)={{1}\over{16\pi^2}}[\Lambda^2+
\omega^2\ln{ {\omega^2}\over{\Lambda^2+\omega^2}}]
\end{equation}
and
\begin{displaymath}
I(x,x)={{\Lambda^2\omega^2}\over{32\pi^2}}+
{{1}\over{64\pi^2}}
\omega^4[\ln{ {\omega^2}\over{\Lambda^2+\omega^2}}-{{1}\over{2}}]+
{{\Lambda^4}\over{64\pi^2}}R(q^2)+{C}
\end{displaymath}
where $R$ denotes the ultraviolet finite expression 
($q^2={{\omega^2}\over{\Lambda^2}}$)
\begin{displaymath}
R(q^2)=\ln (1+q^2)-q^2+{{1}\over{2}}q^4
\end{displaymath}
and $C$ is an $\omega$-independent constant.

In the class of Gaussian states, the integral equation
in Eq.~(\ref{eq:a1}) becomes a simple algebraic equation in 
the mass parameter $\omega$ for any given value of 
$\phi_B$
\begin{equation}
\label{eq:a6}
\omega^2=r_0+3\lambda_0\phi^2_B+3\lambda_0 G(x,x) \,.
\end{equation}
 By introducing dimensionless variables 
\begin{equation}
\label{eq:a7}
               s={{3\lambda_0}\over{16\pi^2}}
\end{equation}
\begin{displaymath}
            r_0=-s\Lambda^2+\epsilon\Lambda^2
\end{displaymath}
\begin{displaymath}
             \phi_B=f\Lambda
\end{displaymath}
 one obtains 
\begin{equation}
\label{eq:a8}
 q^2=\epsilon+3\lambda_0f^2+s q^2\ln{{q^2}\over{1+q^2}} \,.
\end{equation}
The Gaussian effective potential at its minimum, where one replaces 
$q^2=q^2(f^2)$ from Eq.~(\ref{eq:a8}), becomes
\begin{displaymath}
V_{\rm eff}(\phi_B)=V_G(\phi_B,\omega(\phi^2_B))=\Lambda^4u_{\epsilon}(f^2)
\end{displaymath}
where $u_{\epsilon}(f^2)$ can be put in the particularly simple form (up to an 
uninteresting constant term)
%
\begin{equation}
\label{eq:a9}
u_{\epsilon}(f^2)  =   -\frac{q^4(f^2)}{64\pi^2} 
\left[ \ln\frac{q^2(f^2)}{1+q^2(f^2)} + \frac{1}{2} \right] 
+ 
\frac{R\left(q^2(f^2)\right)}{64\pi^2} + \frac{q^4(f^2)}{12\lambda_0} -
\frac{\lambda_0 f^4}{2}
\end{equation}
($u_{\epsilon}(f^2)$ is bounded from below in the limit of large $f^2$).
The meaning of the variational parameter $\omega$ as the shifted-field
mass is further confirmed by 
computing the energy of the one-particle states \cite{stevenson}
\begin{displaymath}
|1\vec p\rangle=a^+_{\omega}(\vec p)|\Psi\rangle_G
\end{displaymath}
where, by using Eq.~(\ref{eq:a6}), one finds \cite{stevenson}
\begin{displaymath}
E_1(\phi_B,\omega(\phi^2_B);\vec p)-
E_G(\phi_B,\omega(\phi^2_B))=\sqrt{ \vec p^2+\omega^2(\phi^2_B)}
\end{displaymath}
so that, indeed, $\omega(\phi^2_B)$ is the gap in the energy
spectrum of the cutoff theory. 
In particular, 
$\omega(0)$ represents the
Gaussian approximation result for $m(r_0,\lambda_0)$ introduced in 
Eq.~(\ref{eq:34}) of Sect.~2.

The qualitative behaviour of $u_{\epsilon}(f^2)$ 
is the following. Let us first
consider the
gap equation at $f=0$. In this case we find 
\begin{displaymath}
q^2(0)=\frac{\epsilon}{1+ s\ln((1+q^2(0))/(q^2(0))} <
\epsilon
\end{displaymath}
which is a positive-definite quantity for any $\epsilon>0$ and one has
\begin{displaymath}
\left.   \frac{d^2u_{\epsilon}}{df^2} \right|_{f=0} = q^2(0) \,.
\end{displaymath}
At large values
of $\epsilon$ the function $u_{\epsilon}(f^2)$ is convex downward and has only
 one minimum
at $f=0$. As $\epsilon$ becomes
smaller
(approaching the regime $\epsilon\sim \exp[-1/s]$) the function 
$u_{\epsilon}(f^2)$
develops secondary maxima and minima since the fourth-derivative at the origin
($F(q^2)=\ln((1+q^2)/q^2)-1/(1+q^2)$)
\begin{displaymath}
\left.   \frac{d^4u_{\epsilon}}{df^4} \right|_{f=0} = 6\lambda_0
\frac{1-sF(q^2(0))}{ 1+ (s/2) F(q^2(0)) }
\end{displaymath}
becomes negative as anticipated in connection with the sign of the coefficient $b$
in Eq.~(\ref{eq:35}) of Sect.~2.
 The secondary minima are located
at non zero values of $f=\pm \bar f$ where 
\begin{displaymath}
          q^2({\bar f}^2)=2\lambda_0{\bar f}^2 \,.
\end{displaymath}
At
\begin{displaymath}
\epsilon=\epsilon_s \sim  \frac{0.8}{e} \, s \,  
\exp \left[ -\frac{1}{2s} \right]
\sim 0.8 \, \epsilon_{\rm max}
\end{displaymath}
($\epsilon_{\rm max}$ being the value above which the effective potential 
has no extrema for $f\neq 0$) one obtains
\begin{displaymath}
              u_{\epsilon_s}({\bar f}^2)=u_{\epsilon_s}(0)  \,.
\end{displaymath}
Thus, in the Gaussian approximation, where $r_c(\lambda_0)=-s\Lambda^2$
(see Eqs.~(\ref{eq:a5}-\ref{eq:a7}) one finds
\begin{displaymath}
r_s(\lambda_0)=r_c(\lambda_0)+\epsilon_s\Lambda^2>r_c(\lambda_0)
\end{displaymath}
as anticipated in Sect.~2. 
For $\epsilon<\epsilon_s$ the minima of $u_{\epsilon}(f^2)$
at $\pm \bar f$ become deeper than its value at $f=0$ and for $\epsilon=0$
one reaches the massless regime, well within the broken phase. In
Fig.~3 
we show the shape of $u_{\epsilon}(f^2)-u_{\epsilon}(0)$ 
for three values of $\epsilon$ indicative 
of the overall situation. 
%
%
%
\begin{figure}[t]
\epsfig{file=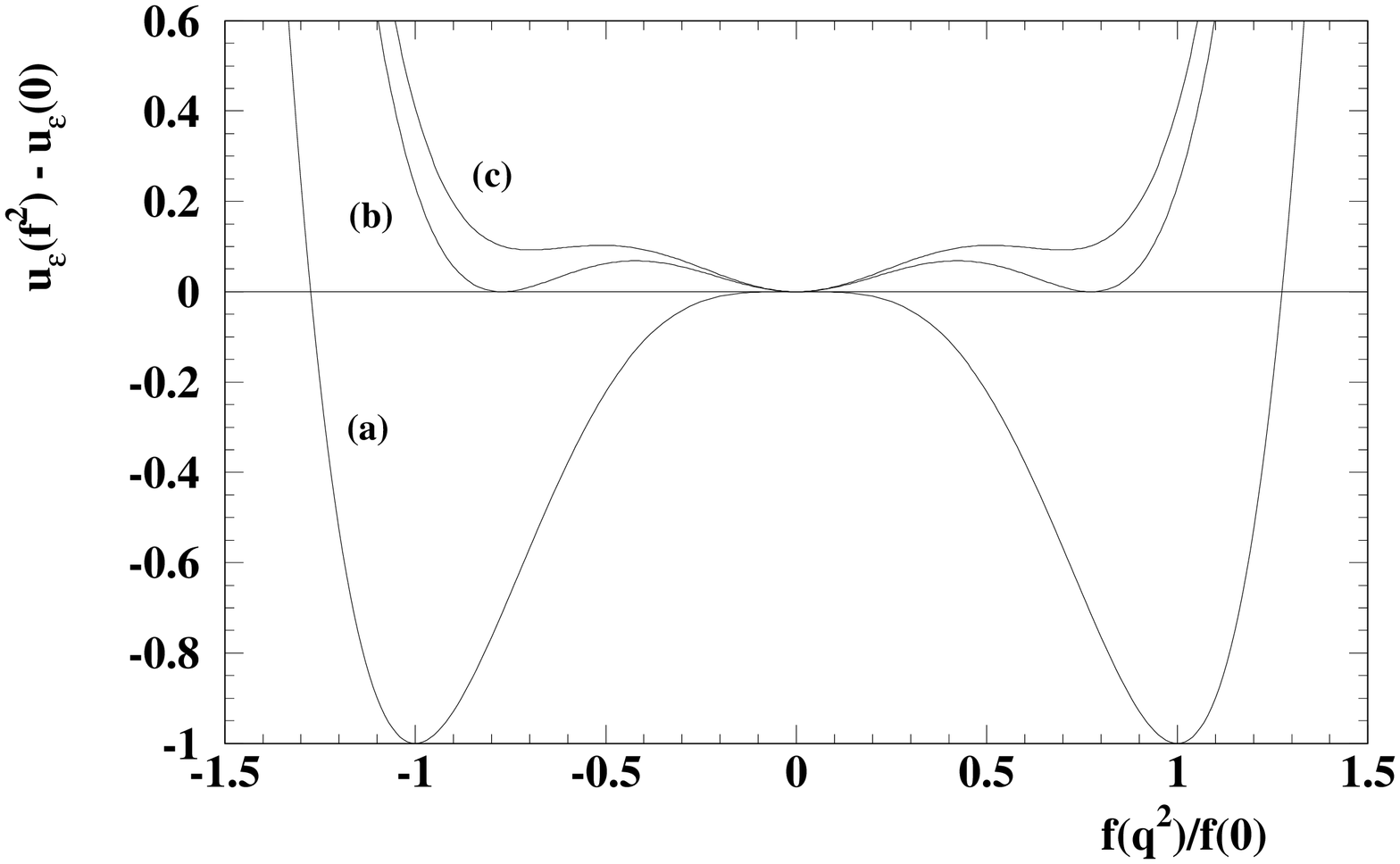,width=\textwidth,height=12truecm}
\caption{$u_{\epsilon}(f^2)-u_{\epsilon}(0)$ versus $f(q^2)/f(0)$ for
three values of $\epsilon$: (a) $\epsilon=0$, (b) $\epsilon=\epsilon_s$,
(c) $\epsilon=0.95 \times \epsilon_{\mathrm{max}}$.}
\label{fig:fig3}
\end{figure}
%
%
%

The special case $\epsilon=0$ is particularly interesting as discussed in 
Sect.~1. In this case, we find the simple relation
for the effective potential at the absolute minima 
$\phi_B=\pm v_B=\pm \bar f \Lambda$
\begin{displaymath}
 W= V_{\rm eff}(\pm v_B)-V_{\rm eff}(0)=-{{m^4_h}\over{128\pi^2}}
\end{displaymath}
with
\begin{equation}
\label{eq:a10}
m^2_h=\omega^2(v^2_B)=2\lambda_0v^2_B=
\Lambda^2\exp \left[ -\frac{8\pi^2}{3\lambda_0} \right] \,.
\end{equation}
To compare with the analogous 1-loop relations, one simply
has to replace $\lambda_0 \to {{3}\over{2}} \lambda_0$ everywhere in
Eq.~(\ref{eq:a10}).

In the relevant region of weak bare coupling, 
where $s<<1$, $r_s$ and $r_c$ are numerically so close 
that a ``direct'' numerical test of the phase transition is not possible
(analogously to the
tiny effect predicted in ref.[41]). However, reliable 
informations can be obtained
by comparing the lattice data with various
models of the effective potential as shown in Sect.~3.

\par For $r_0\neq r_c$, the parametrization of the effective potential
in terms of a renormalized vacuum field $\phi_R$ such that
\begin{displaymath}
\left. \frac{d^2V_{\rm eff}}{d\phi^2_R} \right|_{\phi_R=\pm v_R}=m^2_h 
\end{displaymath}
leads to the more general
relations \cite{zeit} 
(up to terms which vanish in the limit $\Lambda\to\infty$, $\lambda_0\to 0$ 
and $m_h={\mathit{fixed}}$)
%
\begin{equation}
\label{eq:a11}
V_{\rm eff}(\phi_R)
  =  \pi^2\zeta^2\phi^4_R \left( \ln{{\phi^2_R}\over{v^2_R}}-
{{1}\over{2}} \right)  
    +  {{1}\over{4}}(\zeta-1)m^2_h\phi^2_R
(1-{{\phi^2_R}\over{2v^2_R}})
\end{equation}
$\zeta$ being defined through 
\begin{equation}
\label{eq:a12}
                 m^2_h=8\pi^2\zeta v^2_R   \,.
\end{equation}
For all positive values of $\zeta$ the values $\phi_R=\pm v_R$ are minima 
of the effective potential. The minimum has a lower energy than the origin
$\phi_R=0$ if $\zeta<2$. At $\zeta=2$, corresponding to the value of the
bare mass $r_0=r_s$ discussed above,
there is a phase transition to the
broken symmetry phase. Finally, for $\zeta=1$, corresponding to $r_0=r_c$,
one finds the Coleman-Weinberg 
regime \cite{CW} and Eqs.~(\ref{eq:a11},\ref{eq:a12})
reduce to Eqs.~(\ref{eq:20},\ref{eq:21}) of Sect.~1.
The range $0<\zeta<1$, 
corresponding to values $r_0<r_c$
for which the theory cannot be quantized in its symmetric phase, is allowed
by the RG analysis of the effective potential and cannot
be discarded.  For $r_c \leq r_0 \leq r_s$ 
( corresponding to $1 \leq \zeta \leq 2$)
where SSB coexists with
 a physical mass gap in 
the symmetric phase
 $\omega(0)\geq 0$
one finds
\begin{equation}
\label{eq:a13}
{{\omega^2(0)}\over{m^2_h}}<s~{{\zeta-1}\over{\zeta}} \to 0
\end{equation}
in the bare weak-coupling limit where $\lambda_0$ and $s$ vanish at
$m_h={\mathit{fixed}}$. Thus,  the mass gap of the symmetric phase
becomes infinitesimal in units of $m_h$ in agreement  with the general
condition for SSB in Eq.~(\ref{eq:36}) of Sect.~2. This result
confirms the  remarkable consistency of our definition of the
continuum limit.


\begin{thebibliography}{99}
\bibitem{beta} Eq.~(\protect\ref{eq:8}) predicts an unphysical Landau
pole at 1-loop or an ultraviolet fixed point at non-zero bare coupling
at 2-loops. This structure reproduces itself in higher orders
\protect\cite{shirkov}. Indeed, in odd orders 
$\beta^{\mathrm{1-loop}}$, $\beta^{\mathrm{3-loop}}$,
$\beta^{\mathrm{5-loop}}$, each exhibits a Landau singularity whereas in
even orders $\beta^{\mathrm{2-loop}}$, $\beta^{\mathrm{4-loop}}$, have an
ultraviolet fixed point whose magnitude decreases by increasing the
perturbative order. In this situation, for a sufficiently high
ultraviolet cutoff, the conventional approach based on hierarchy of
leading logarithmic, next-to-leading, next-to-next-to-leading, $\dots$
effects looses its meaning and, for any non-vanishing $\lambda_{\mathrm
R}$, no consistent continuum limit is possible in perturbation theory.
\end{thebibliography}
\end{document}